\documentclass[11pt,twoside]{article}

\usepackage{asp2006}
\usepackage{epsf}
\usepackage{psfig}
\usepackage{lscape}

\newcommand{\msun}{\mbox{${\rm M}_{\odot}$}}
\newcommand{\lsun}{\mbox{${\rm L}_{\odot}$}}
\begin{document}
\title{The Relation Between the Globular Cluster Mass and Luminosity Functions}   %%% Fill in title
\author{J.~M.~Diederik Kruijssen$^{1,2}$ and Simon~F.~Portegies~Zwart$^2$}   %%% Fill in author names
\affil{$^1$Astronomical Institute, Utrecht University, PO Box 80000, 3508 TA Utrecht, The Netherlands; {\tt kruijssen@astro.uu.nl}}
\affil{$^2$Leiden Observatory, Leiden University, PO Box 9513, 2300 RA Leiden, The Netherlands}

\begin{abstract} %%% Abstract to run on from here.
The relation between the globular cluster luminosity function (GCLF, ${\rm d}N/{\rm d}\log{L}$) and globular cluster mass function (GCMF, ${\rm d}N/{\rm d}\log{M}$) is considered. Due to low-mass star depletion, dissolving GCs have mass-to-light ($M/L$) ratios that are lower than expected from their metallicities. This has been shown to lead to an $M/L$ ratio that increases with GC mass and luminosity. We model the GCLF and GCMF and show that the power law slopes inherently differ (1.0 versus 0.7, respectively) when accounting for the variability of $M/L$. The observed GCLF is found to be consistent with a Schechter-type initial cluster mass function and a mass-dependent mass-loss rate.
\end{abstract}

%%% MAIN BODY OF TEXT GOES HERE. CONSULT "INSTRUCTIONS FOR AUTHORS USING
%%% LATEX2E MARKUP", SECTIONS 2.3-2.6 FOR HELP WITH EQUATIONS, FIGURES,
%%% AND TABLES.

%\section{}   %%% Top level section head (remove "%" symbol)
%\subsection{}   %%% Second level section head (remove "%" symbol)
%\subsubsection{}   %%% Lowest level section head (remove "%" symbol)
%\section*{}    %%% Unnumbered top level section head (remove "%" symbol)
%\subsection*{}   %%% Unnumbered second level section head (remove "%" symbol)

\section{Introduction}
Even though globular cluster systems (GCSs) are considered to have formed during mergers, the shape of the globular cluster luminosity function (GCLF, ${\rm d}N/{\rm d}\log{L}$) differs fundamentally from that of young massive clusters (YMCs) in merging galaxies. The luminosity function of YMCs follows a power law with index $-2$ over the full mass range down to a few $100~\msun$, while the GCLF peaks at $\sim 10^5~\lsun$. This has been attributed to the ongoing tidal disruption of globular clusters (GCs) and the resulting destruction of low-mass, faint GCs \citep{fall01,vesperini01}. In studies considering this mechanism, the mass-loss rate of GCs is considered to be independent of their mass (corresponding to a disruption time $t_{\rm dis}\propto M$) and the mass-to-light ratio ($M/L$) is assumed to be constant \citep[e.g.,][]{fall01,jordan07}.

\begin{figure}
\center\resizebox{12.4cm}{!}{\plottwo{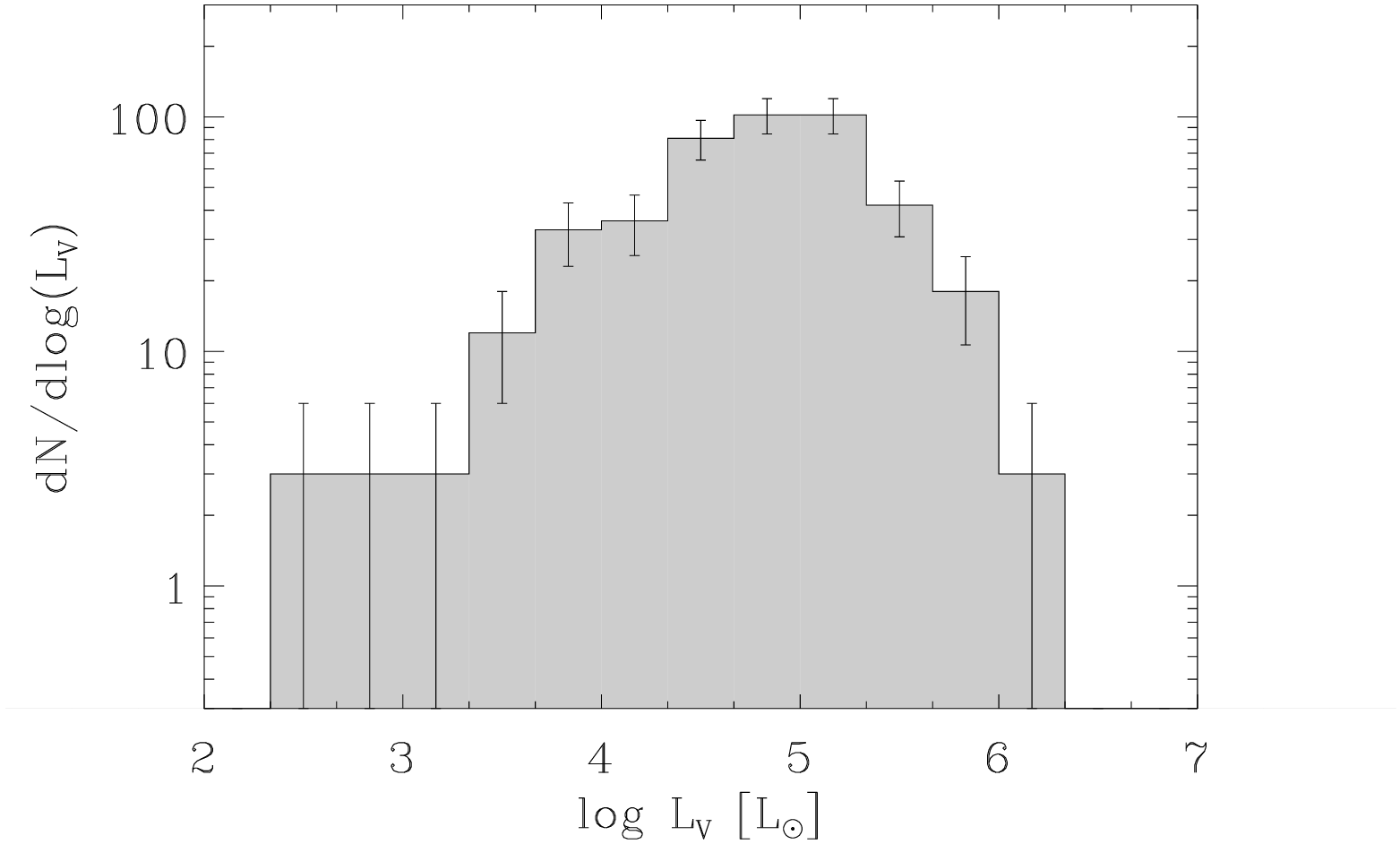}{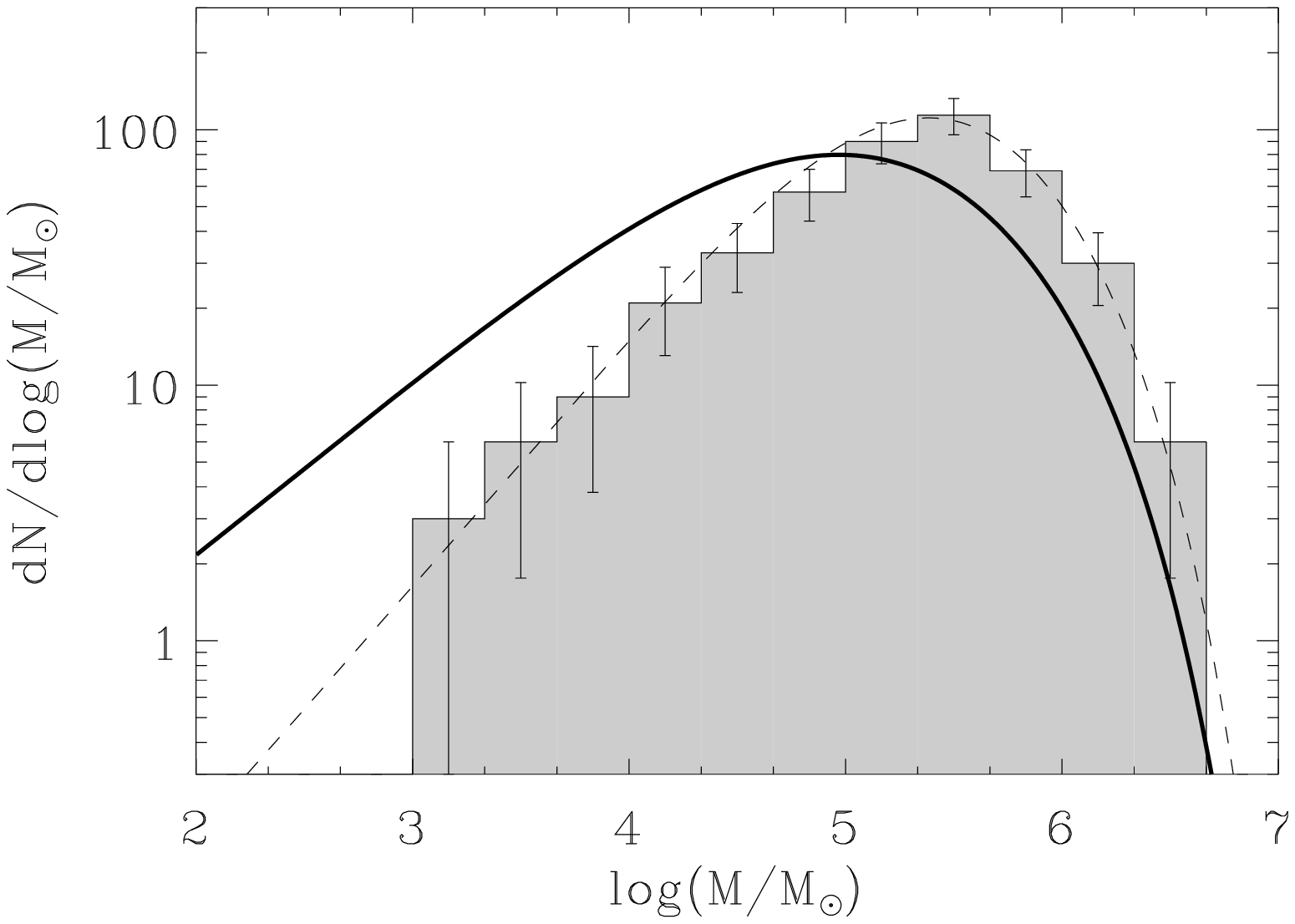}}
\caption[]{\label{fig:histcan}
{\it Left}: The observed GCLF of Galactic GCs \citep{harris96}. {\it Right}: The inferred GCMF of Galactic GCs (histogram) using $M/L_V=3$ \citep[as in][]{fall01}. Overplotted is our model MF with a mass-dependent mass-loss rate (solid line, see Section~\ref{sec:model}) adopting a dissolution timescale $t_0=1.6$~Myr. The dashed line shows the model for a cluster mass-independent mass-loss rate \citep[as in][]{fall01}. Error bars are $1\sigma$ Poissonian.}
\end{figure}

In order to trace back the merger history of galaxies using their globular cluster mass function (GCMF, ${\rm d}N/{\rm d}\log{M}$), it is essential to obtain an accurate description for its evolution. Although the above studies reproduced the peaked shape of the GCMF, they did not account for two aspects of GC evolution:
\begin{itemize}
\item[(1)]
The mass-loss rate of GCs does depend on cluster mass \citep{baumgardt01,baumgardt03,lamers05,larsen09}, corresponding to $t_{\rm dis}\propto M^\gamma$ with $\gamma\sim 0.7$ (see Eq.~\ref{eq:dmdt}). This is due to the nonlinear scaling of the disruption time with the half-mass relaxation time \citep[$t_{\rm dis}\propto t_{\rm rh}^{0.75}$,][]{portegieszwart98,fukushige00}. A lower value of $\gamma$ means that the dissolution rate of low-mass clusters is slowed down with respect to massive clusters.
\item[(2)]
The $M/L$ ratio of GCs is not constant \citep{mandushev91,baumgardt03,kruijssen08,kruijssen09} due to the preferential ejection of faint, low-mass stars from dissolving GCs. Because low-mass GCs have on average lost a larger fraction of their initial masses, $M/L$ increases with mass or luminosity.
\end{itemize}
The effect of these aspects of GC evolution on the relation between the GCLF and the GCMF has been considered by \citet{kruijssen09b}. The slope of the disruption-dominated low-mass side of the GCMF is always equal to $\gamma$ \citep{fall01,lamers05}. A mass-dependent mass-loss rate ($\gamma=0.7$) therefore yields a different GCMF slope than a mass-independent mass-loss rate ($\gamma=1$). The observed slope of the GCLF $\sim 1$, which is seemingly consistent with the latter if a constant $M/L$ ratio is assumed (see Fig.~\ref{fig:histcan}). However, the trend of increasing $M/L$ ratio with luminosity implies that this is not necessarily true because the slopes of the GCLF and the GCMF fundamentally differ. \citet{kruijssen09b} have shown that the observed GCLF is in fact consistent with a GCMF with a low-mass slope of $\sim 0.7$ when accounting for the trend of $M/L$ ratio with luminosity.

We revisit the calculations of \citet{kruijssen09b} and include a more detailed model for the evolution of the stellar mass function (SMF) within a star cluster. Previously, the low-mass star depletion was approximated by increasing the lower stellar mass limit of the SMF. In reality, the SMF evolves more gradually. This has been included in new models of the evolution of the SMF in dissolving star clusters \citep{kruijssen09c}, which are used here.

\section{Star cluster models} \label{sec:model}
We use the parameterised cluster model {\tt SPACE} \citep{kruijssen08b,kruijssen09c}, which incorporates the effects of stellar evolution, stellar remnant production, dynamical dissolution and energy equipartition. The mass-loss rate due to dissolution follows from $t_{\rm dis}=t_0 M^\gamma$ as \citep{lamers05}:
\begin{equation}
\label{eq:dmdt}
\left(\frac{{\rm d}M}{{\rm d}t}\right)_{\rm dis}=-\frac{M}{t_{\rm dis}}=-\frac{M^{1-\gamma}}{t_0} ,
\end{equation}
with $t_0$ a constant that represents the rapidity of dissolution and depends on the environment.

The evolution of the SMF is computed by considering the ejection rate as a function of stellar mass \citep{kruijssen09c}. The adopted method accounts for mass segregation and dissolution in a tidal field by using the timescale on which energy equipartition is reached for different stellar masses and by comparing the stellar velocities with the escape velocity.

\section{The inferred GCMF using realistic mass-to-light ratios}
\begin{figure}
\center\resizebox{12.4cm}{!}{\plottwo{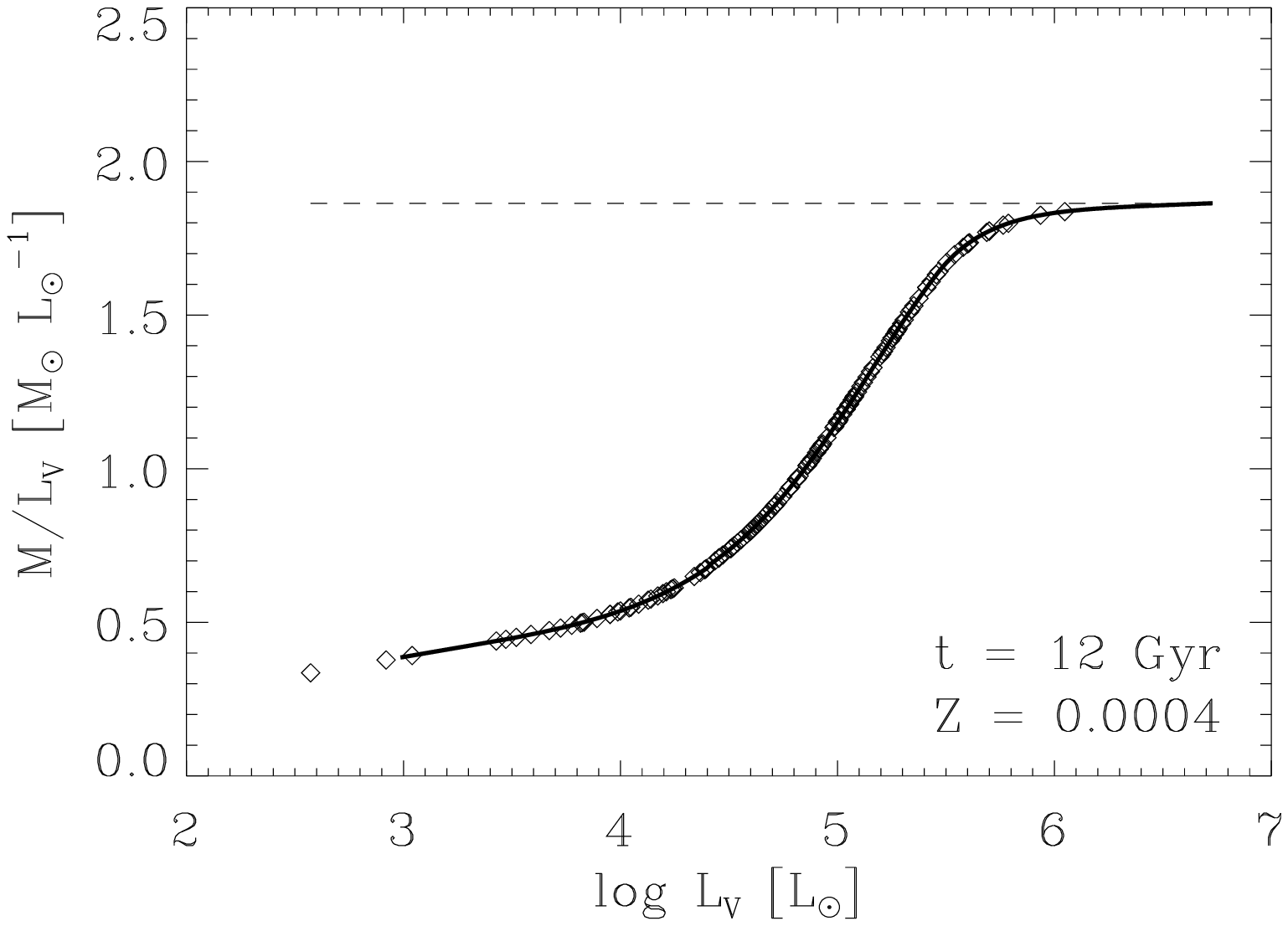}{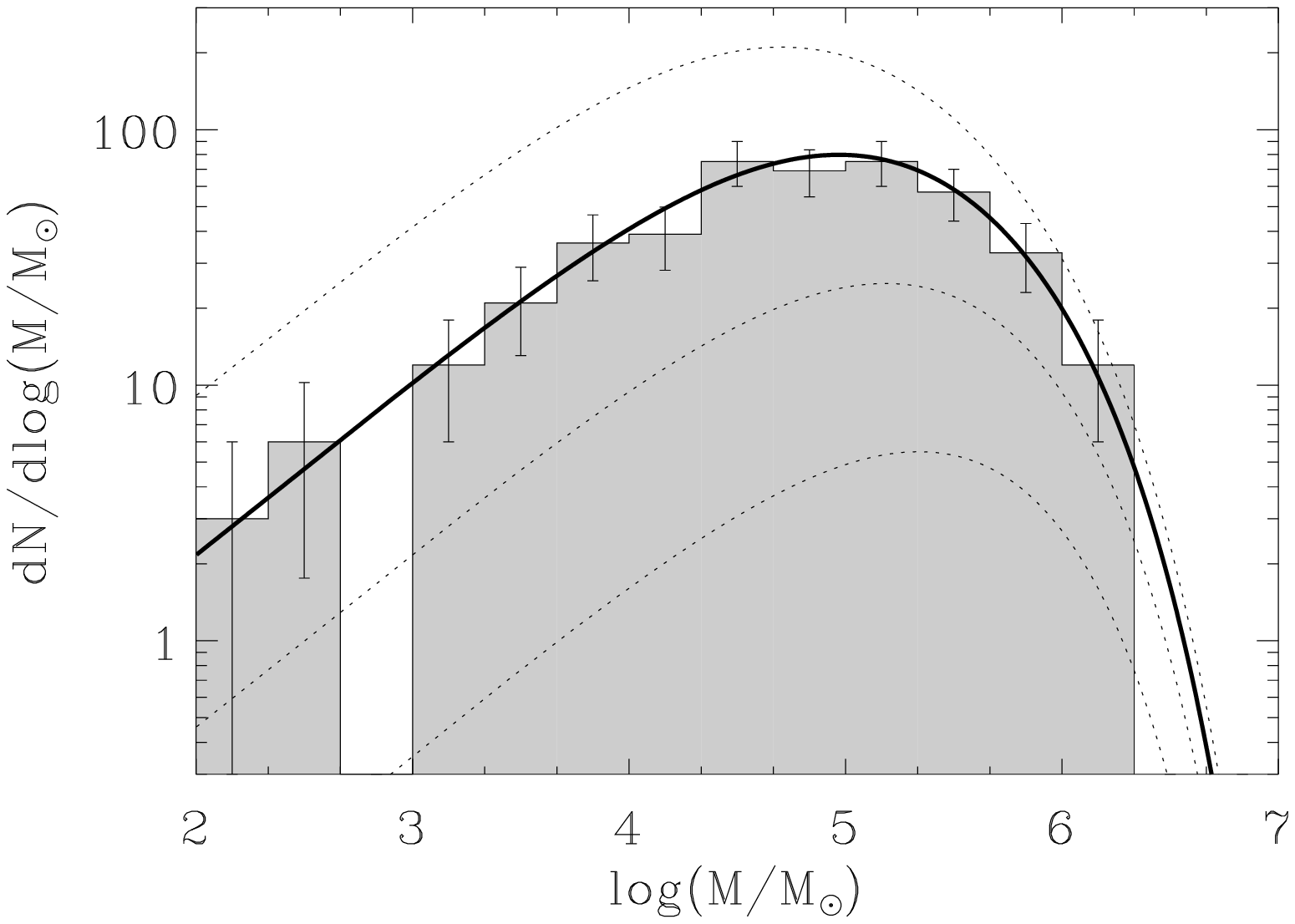}}
\caption[]{\label{fig:histnew}
{\it Left}: Modeled relation between $M/L_V$ ratio and luminosity $L_V$ (solid line). Diamonds mark the luminosities of observed Galactic GCs, and the dashed line denotes the constant $M/L_V$ ratio that would be obtained if low-mass star depletion were neglected. {\it Right}: GCMF derived from the GCLF using a luminosity-dependent $M/L_V$ ratio. The solid curve is the same as in Fig.~\ref{fig:histcan} while the dotted curves represent models for (from bottom to top) $\log{(t_0/{\rm Myr})}=\log{1.6}+\{-0.5,-0.25,0.25\}$. Error bars are $1\sigma$ Poissonian.}
\end{figure}
The modeled relation between the $M/L_V$ ratio and $V$-band luminosity $L_V$ is shown in the left-hand panel of Fig.~\ref{fig:histnew}. Variations in dissolution timescale and metallicity move this relation horizontally and vertically \citep{kruijssen08}. When considering a real GC system, this induces scatter off the depicted relation. \citet{kruijssen09b} considered this scatter in a detailed model of the Galactic GC system and concluded that a relation such as the one that is shown in Fig.~\ref{fig:histnew} can still be used as a mean for the entire GC population.

The GCMF that is derived from the observed Galactic GCLF using the relation between $L_V$ and $M/L_V$ is shown in the right-hand panel of Fig.~\ref{fig:histnew}. It is compared to our modeled GCMF that has evolved from a \citet{schechter76} initial cluster mass function with index -2 and exponential cut-off mass $M_*=2.5\times 10^6~\msun$. The modeled and observed GCMFs are fully consistent, implying that the mass-loss rate of GCs is indeed mass-dependent ($t_{\rm dis}\propto M^\gamma$, with the proportionality constant depending on the environment) and that there indeed exists a weak trend of $M/L$ ratio with mass and luminosity. As such, the results shown in Fig.~\ref{fig:histnew} confirm the conclusions of \citet{kruijssen09b} with more detailed cluster models.

\section{Discussion}
Using a realistic model for the evolution of the SMF in GCs, we have shown that the slopes of the GCLF and the GCMF differ. As GCs dissolve, their masses decrease more rapidly than their luminosities due to the ejection of faint, low-mass stars. As a result, the slope of the GCLF is $\sim 1$ for luminosities below the peak luminosity, while the slope of the GCMF is $\sim 0.7$ for masses below the peak mass. This is consistent with a mass-loss rate that depends on cluster mass and on the environment.

The new cluster models \citep{kruijssen09c}, in which we account for the changing slope of the stellar mass function rather than shifting the lower stellar mass limit, show that the results from \citet{kruijssen09b} also hold when more detailed models are applied. This substantiates that care should be taken when comparing GCLFs and GCMFs. Because of the relatively low number of GCs for which dynamical masses have been determined, we argue that the best method to compare them would be to model the GCLF rather than the GCMF, while accounting for the variability of the $M/L$ ratio. This would allow for a more accurate interpretation of the GCLF when studying correlations between its properties and the history of its galaxy.

\acknowledgements %%% Text of acknowledgements runs on after this command.
JMDK thanks East Tennessee State University for hosting an excellent conference and for financial support. The Leids Kerkhoven-Bosscha Fonds (LKBF) is acknowledged for supporting attendance to the conference. This research is supported by the Netherlands Advanced School for Astronomy (NOVA), the 
LKBF and the Netherlands Organisation for Scientic Research (NWO), grant numbers 021.001.038, 639.073.803, and 643.200.503.

%%% THE BIBLIOGRAPHY
%%%
%%% CONSULT SECTION 3 OF "INSTRUCTIONS FOR AUTHORS" FOR HOW TO USE NATBIB.
%%% AUTHORS ARE ENCOURAGED TO USE EITHER THE "THEBIBLIOGRAPY" ENVIRONMENT
%%% BY UNCOMMENTING (DELETING THE "%" SYMBOL) THE COMMANDS BELOW, OR BY
%%% USING THE BIBTEX ENVIRONMENT. TO FIND OUT WHICH IS APPLICABLE TO YOUR
%%% CONTRIBUTION, CONSULT THE VOLUME EDITORS FOR YOUR PROCEEDINGS.
%%%

%\bibliographystyle{apj}
%\bibliography{mybib}

\end{document}